\documentclass[prb,twocolumn,showpacs,amsmath,amssymb,superscriptaddress]{revtex4-2}

\usepackage{graphicx}
\usepackage{ifthen}
\usepackage{color}
\usepackage{bm}
\usepackage{braket}
\usepackage{multirow}
\usepackage{hyperref}
\hypersetup{backref=true,
 pdfnewwindow=true, colorlinks=true,
 linkcolor=blue, anchorcolor=blue,
 citecolor=blue, filecolor=blue,
 menucolor=blue, urlcolor=blue}

\usepackage{pdfpages}
\usepackage{pgffor}
\makeatletter
\AtBeginDocument{\let\LS@rot\@undefined}
\makeatother

\begin{document}

\title{Orbital magnetization of a metal is not a bulk property in the mesoscopic regime}

\author{Kevin Moseni}
\affiliation{Materials Science and Engineering, University of California Riverside, Riverside, CA
  92521, USA}
\author{Sinisa Coh}
\affiliation{Materials Science and Engineering, University of California Riverside, Riverside, CA
  92521, USA}
\affiliation{Mechanical Engineering, University of California Riverside, Riverside, CA
  92521, USA}

\date{\today}

\begin{abstract}
We find that, in the mesoscopic regime, modification of the material's surface can induce an extensive change of the material's magnetic moment.  In other words, perturbation of order $N^2$ atoms on the surface of a 3-dimensional solid can change the magnetic moment proportionally to $N^3$.  When the solid's surface is perturbed, it triggers two changes in the magnetization. One arises from variations of the electron wavefunction and energy, while the other arises from a modification in the kinetic angular momentum operator. In the macroscopic regime of our model, these two bulk effects cancel each other, resulting in no impact of the surface perturbation on the magnetization --- consistent with prior work. 
In the mesoscopic regime, we find a departure from this behavior, as the cancelation of two terms is not complete.
\end{abstract}

\maketitle

\section{Introduction} 
In a ferromagnet, the magnetic moment arises primarily from the unequal population of electrons with different spin states. A smaller, but significant contribution, known as orbital magnetization, originates from the microscopic spatial motion of electrons throughout the material. Some of these microscopic orbital electron currents flow around individual atoms in the bulk, while other currents traverse the surface of the sample, as demonstrated in Ref.~\onlinecite{Thonhauser2005} using a framework of localized Wannier states.  Although only a fraction of electrons participate in surface currents, their collective effect contributes to the magnetic dipole moment, scaling with the volume of the sample (area, in two dimensions).

The question then arises whether the magnetic moment of the ferromagnet could be modified by perturbing the surface of the material? For instance, one may wonder if adsorbing atoms to the surface of a solid could induce currents and consequently change the magnetic dipole of the solid, in proportion to the volume of the solid? In other words, we are asking whether perturbing order $N^2$ atoms on the surface of a 3-dimensional solid could change the magnetic moment proportional to $N^3$?  Or, similarly, could perturbation of order $N$ atoms on the {\it edge} of a 2-dimensional solid change the magnetic moment in proportion to $N^2$?

The seminal work from Ref.~\onlinecite{Thonhauser2005} demonstrated that none of these scenarios is possible for insulating systems.  In an insulating system, surface currents are quite remarkably determined by the material properties deep in the bulk of the material.  Intuitively, one would expect such a statement to also extend to metallic cases. Reference~\onlinecite{Ceresoli2006} gives heuristic reasons why magnetization in a metal is equally well determined by the properties of the bulk of the material, as in the case of an insulator.  (The same was also suggested for topological insulators in Refs.~\onlinecite{Ceresoli2006, Chen2012, wang2022}.)  Additional support is given by the semiclassical formulation of orbital magnetization from Ref.~\onlinecite{Di2005} as well as the long-wave perturbation from Ref.~\onlinecite{Shi2007}. A more recent proof that orbital magnetization in a metal is a bulk property relies on a local measure of the orbital moment from Refs.~\onlinecite{Bianco2013, bianco2016, AntimoResta2016, Resta2018}.

\begin{figure}[!t]
\centering
\includegraphics{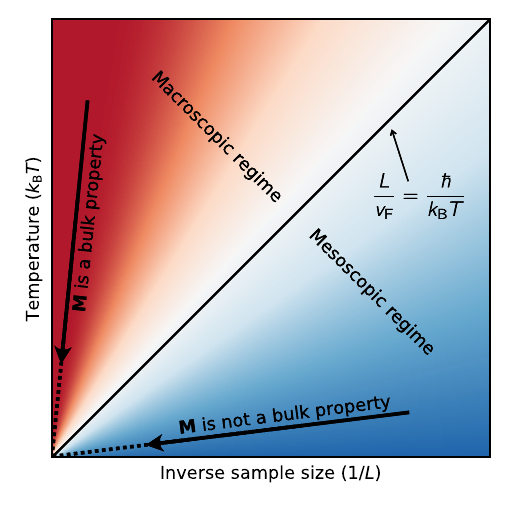}
\caption{\label{fig:phase} Orbital magnetization is not a bulk property in the mesoscopic regime (blue color).  This regime persists in the limit of the infinite sample size (horizontal arrow) as long as the temperature is low enough so that one stays in the mesoscopic regime. The thin diagonal black line separates mesoscopic from the macroscopic regime, as given by Eq.~\ref{eq:meso} in the text.  If the limit to infinite sample size is taken along a vertical trajectory shown in the red region, the orbital magnetization is a bulk property, consistent with previous works.~\cite{Di2005, Thonhauser2005, Ceresoli2006, Shi2007, Chen2012, Bianco2013, bianco2016, AntimoResta2016, wang2022, Resta2018}}
\end{figure} 

In this paper, our focus lies on a distinct range of length and temperature scales, one that complements the scope of previous investigations. Previous studies can be applied to the macroscopic regime, which we define as 
\begin{align}
\frac{L}{v_{\rm F}}  \gg \frac{\hbar}{k_{\rm B} T}.
\label{eq:macro}
\end{align}
Here $v_{\rm F}$ is the electron's Fermi velocity and $L$ is a length of the sample.  In other words, in the macroscopic regime, the time of flight of the electron across the sample ($L/v_{\rm F}$) exceeds the time scale associated with the thermal energy, $\hbar / (k_{\rm B} T)$.  In the macroscopic regime, our findings corroborate the conclusions drawn in Refs.~\onlinecite{Di2005, Thonhauser2005, Ceresoli2006, Shi2007, Chen2012, Bianco2013, bianco2016, AntimoResta2016, wang2022, Resta2018}. Specifically, the surface modifications do not lead to extensive changes in magnetization.  Nevertheless, an intriguing situation emerges when we shift to the opposite regime,
\begin{align}
 \frac{L}{v_{\rm F}} < \frac{\hbar}{k_{\rm B} T},
\label{eq:meso}
\end{align}
which we refer to as the mesoscopic regime.  Our work shows that in the mesoscopic regime the surface can indeed change the overall magnetic moment of the sample, in proportion to the volume of the sample. \footnote{Strictly speaking, in the mesoscopic regime we need to require that additionally $k_{\rm B} T$ is larger than the typical level spacing (scaling as $1/L^2$).  If $k_{\rm B} T$ is smaller than the level spacing, the model is in the so-called microscopic regime.  We refer the reader to Ref.~\onlinecite{gurevich1997orbital} where these limits are studied in detail for the related case of Landau diamagnetism. For the present work, the distinction between microscopic and mesoscopic regimes is not relevant.}  Figure~\ref{fig:phase} shows the macroscopic and mesoscopic regimes as a function of sample size and temperature. 

We stress that in this work the macroscopic regime does not simply correspond to the $L \rightarrow \infty$ limit of infinite sample size.  From the definition of the macroscopic regime (Eq.~\ref{eq:macro}) and the mesoscopic regime (Eq.~\ref{eq:meso}) it is clear that one can reach the limit of an infinite sample size ($L \rightarrow \infty$) in both the macroscopic and mesoscopic cases, depending on how quickly $T \rightarrow 0$ relative to $L \rightarrow \infty$.  Two qualitatively different limits are indicated by vertical and horizontal arrows in Fig.~\ref{fig:phase}.  Along the vertical arrow, in the red region, magnetization is a bulk property, but along the horizontal arrow, in the blue region, it is not a bulk property.

This paper is structured as follows. In Sec.~\ref{sec:motivation}, we motivate the model and provide a general procedure for modifying the edge (surface) of a model so that it induces extensive orbital magnetization. In Sec.~\ref{sec:construction}, we provide a concrete example in the form of a two-dimensional tight-binding model. In Sec.~\ref{sec:results}, we show our main results. In Sec.~\ref{sec:conclusions}, we summarize and provide an outlook.

\section{Motivation}\label{sec:motivation}

Before introducing our numerical model, we first motivate it by considering a continuous one-particle effective Hamiltonian, denoted $H^0_{\rm c}$, for a periodic infinite solid. For simplicity we work in two dimensions, but generalization to higher dimensions is straightforward by stacking lower-dimensional models. When dealing with the two-dimensional models, we will refer to the boundary as {\it edge} instead of {\it surface}, which we reserve for three-dimensional solids. To simplify our analysis, throughout this work we neglect spin, self-consistency, many-electron effects, and disorder.  Our system is assumed to be in thermal equilibrium.  We ignore any temperature effects beyond electron occupation smearing.

To motivate our construction, we recall first that the complete basis of the eigenstates of $H^0_{\rm c}$ can be expressed in the Bloch form, $\psi_{\bm k} ({\bm r}) = e^{i {\bm k} \cdot {\bm r}} u_{\bm k} ({\bm r})$. However, not every eigenstate of $H^0_{\rm c}$ has the Bloch form. Generally, we can construct arbitrary linear combinations of states that share the same eigenvalue $E_{\bm k} = E$, and the resulting function
\begin{align}
\phi_{E} ({\bm r}) = \int_0^1 e^{i f(s)} \psi_{\bm{k} (s)} ({\bm r}) ds
\end{align}
is a valid eigenstate of $H^0_{\rm c}$.  Here
$$s \rightarrow {\bm k} (s)$$ 
is a continuous parameterization of a curve in the Brillouin zone along which energy is constant, 
$$E_{{\bm k}(s)} = E.$$

For now we limit $f(s)$ so that it is periodic, 
$$f(0) = f(1).$$
We choose $f(s)$ so that $\phi_{E} ({\bm r})$ is as localized as possible in real space. $\phi_{E}$ is only algebraically localized due to integration over part of the Brillouin zone, unlike exponential localization of a Wannier function.  Another difference to the Wannier function is that $|\phi_{E} ({\bm r})|^2$ remains stationary in time, just like a Bloch state, in contrast to the Wannier function that disperses in space during its time evolution.

By selecting a fixed $f(s)$, next we create a family of functions, $\phi_{m E}$, for any integer $m$, defined as follows,
\begin{align}
\phi_{m E} ({\bm r}) = \int_0^1 e^{i 2 \pi m s} e^{i f(s)} \psi_{\bm{k} (s)} ({\bm r}) ds.
\label{eq:transf_m}
\end{align}
Note, trivially, that $$\braket{\phi_{m E} | \phi_{m' E'}} = \delta_{m m'} \delta_{E E'}.$$ Therefore, $\phi_{m E}$ for all $m$ and $E$ span the same vector space as the Bloch states. The transformation defined in Eq.~\ref{eq:transf_m} for any integer $m$ therefore has parallels to a shift of a Wannier function by a lattice vector $\bm R$, as a set of shifted Wannier functions spans the same vector space of Bloch states.

Let us now take the simplest case of $H^0_{\rm c}$ corresponding to the free-electron system with mass $m_{\rm e}$. In this case $\phi_{mE}({\bm r})$ in cylindrical coordinates is simply $$\phi_{mE}({\bm r}) \sim e^{i m \varphi} J_m \left(\frac{\sqrt{2 m_{\rm e} E}}{\hbar} r \right)$$ where $J_m$ is the Bessel function of the first kind.

Trivially, the expectation value of the angular momentum operator $L_z$ is
\begin{align}
\bra{\phi_{m E}} L_z \ket{\phi_{m E}} = \hbar m.
\label{eq:phi_Lz}
\end{align}
Therefore, each state $\phi_{m E}$ carries angular momentum $\hbar m$, and orbital magnetic moment $\mu_{\rm B} m$.  Let us now confine our system to a circular region with radius $R$.  From elementary properties of Bessel functions it follows that states with large enough $m$, close to $R \frac{\sqrt{2 m_{\rm e} E}}{\hbar}$, are localized near the edge of the sample ($r \approx R$).  These states carry an angular momentum $\hbar m \sim R^1$, and the number of these states also scales as $\sim R^1$.  Therefore, one might hope that tweaking the electron potential $V^{\rm edge}$ near the edge of the sample could modify these states and induce a net orbital moment that scales as $\sim R^2$. Specifically, if one could construct an edge potential $V^{\rm edge}$ satisfying
\begin{align}
\bra{\phi_{m E}} V^{\rm edge} \ket{\phi_{m' E}} \sim m \delta_{m m'}
\label{eq:cont_m}
\end{align}
then this would be a good candidate edge perturbation, as it breaks the time-reversal symmetry by acting differently on states with different $m$.  For example, one of the effects of this perturbation would be to push $m<0$ states below the Fermi level, and $m>0$ states above the Fermi level, thus inducing a net magnetic dipole. However, as we discuss later, in Sec.~\ref{subsec:polarization}, there are other changes to the magnetic moment induced by the edge perturbation, such as changes to the electron wavefunction, as well as changes to the angular momentum operator itself.

\section{Model construction}\label{sec:construction}

With this motivation, we now set out to create edge potential satisfying Eq.~\ref{eq:cont_m} in a concrete finite-size model. For numerical convenience we use a tight-binding approach. 

\begin{figure}[!t]
\centering
\includegraphics{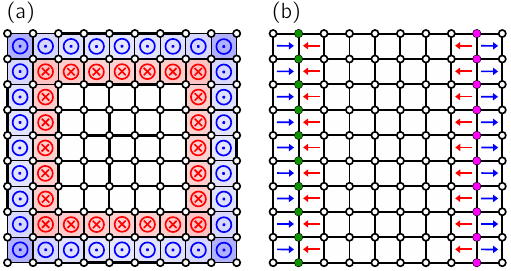}
\caption{\label{fig:mod} (a) Perturbation $V^{\rm edge}$, given by Eq.~\ref{eq:V_edge}, induces complex phases on hopping elements near the edge.  Blue and red colors represent different signs of effective local {\it magnetic} field on the edge. (b) Perturbation $V'^{\rm edge}$, given by Eq.~\ref{eq:vprime}, changes the onsite energies of green and purple orbitals on the left and right edges.  Arrows represent directions of effective local {\it electric} field at the edge.  In both cases (a) and (b) magnitude of effective magnetic and electric fields on the edge is independent of $N$. (Effective internal magnetic and electric fields are generically present in materials with broken time-reversal or inversion symmetry, and they do not require application of external fields.)}
\end{figure}

To construct the tight-binding model, we project our continuous free-electron Hamiltonian $H^0_{\rm c}$ on the basis of a $N \times N$ square mesh of s-like orbitals separated from each other by a distance $a$ (orbitals are sketched as black circles in Fig.~\ref{fig:mod}). We label the orbital at site $i$ as $\ket{i}$. For the position operators $x$ and $y$, we assume that they are diagonal,
\begin{align}
\bra{i} x \ket{j} & = x_i \delta_{ij} \\
\bra{i} y \ket{j} & = y_i \delta_{ij}.
\end{align}
For convenience, we work with the centered operators, defined as 
\begin{align}
\tilde{x} &= x - \frac{\sum_i x_i}{N^2}, \\
\tilde{y} &= y - \frac{\sum_i y_i}{N^2}
\end{align}
so that, by construction, the center of mass of $\tilde{x}_i$ and $\tilde{y}_i$ over all orbitals in the model is zero. We also define the following quantity $\tilde{L} (A)$ for any operator $A$,
\begin{align}
\tilde{L} (A) 
&=\frac{i m_{\rm e}}{\hbar} \left( \tilde{x} [A, \tilde{y}] - \tilde{y} [A, \tilde{x}] \right) \notag \\
&=\frac{i m_{\rm e}}{\hbar} \left( \tilde{x} A \tilde{y} - \tilde{y} A \tilde{x} \right).
\end{align}
Clearly, $\tilde{L} (H)$ corresponds to the angular momentum operator for a system described by the Hamiltonian $H$. 

\begin{figure*}[!t]
\centering
\includegraphics{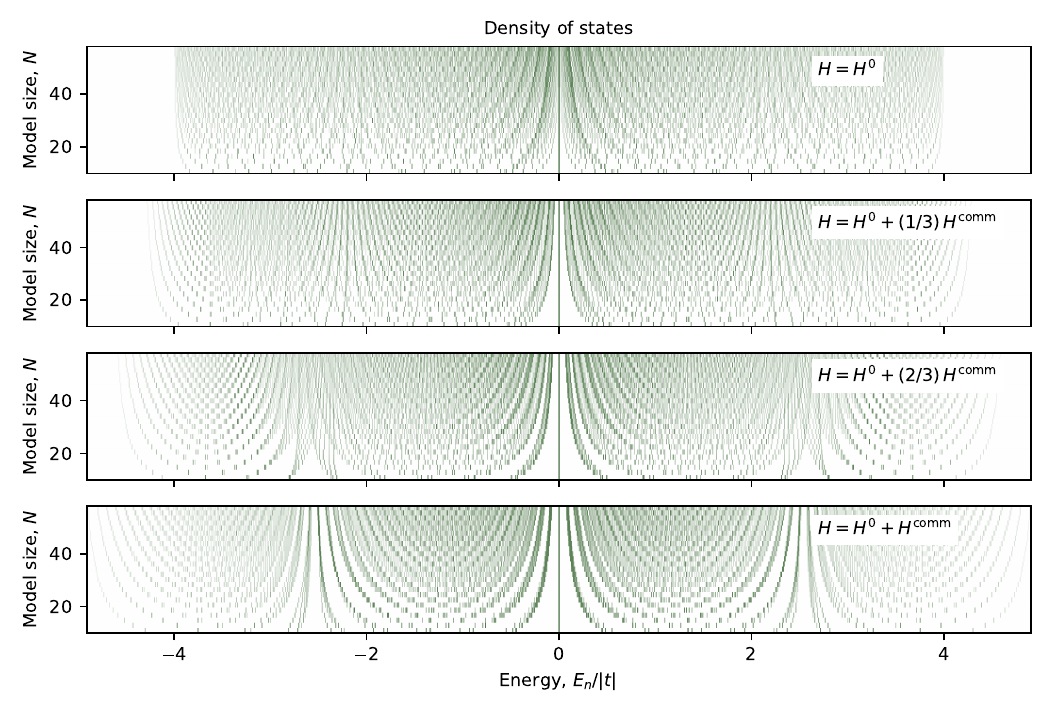}
\caption{\label{fig:dos} Density of states for Hamiltonian $H^0 + \beta H^{\rm comm}$ as a function of $N$.  The parameter $\beta$ ranges from 0 (top panel) to 1 (bottom panel) in steps of $1/3$.  Clearly, the addition of $H^{\rm comm}$ rearranges the spectrum by opening a set of small gaps $\Delta \sim 1/N$.}
\end{figure*}

\subsection{General procedure}

Now we describe a general five-step procedure that constructs edge potential $V^{\rm edge}$ satisfying Eq.~\ref{eq:cont_m} given bulk hamiltonian $H^0$, and computes the magnetic dipole of the sample.
\begin{align}
&{\rm Step \ 1: \ choose \ } H^0.
\notag \\
&{\rm Step \ 2: \ construct \ } H^{\rm comm} {\rm \ from \ } H^0.
\notag \\
&{\rm Step \ 3: \ construct \ } V^{\rm edge} {\rm \ from \ } \tilde{L}(H^0).
\notag \\
&{\rm Step \ 4: \ diagonalize \ } H = H^0 + H^{\rm comm} + V^{\rm edge}.
\notag \\
&{\rm Step \ 5: \ compute \ } m_{\rm dip} = \frac{e}{2 m_{\rm e}} \sum_n \bra{\psi_n} \tilde{L}(H) \ket{\psi_n} f_n.
\notag
\end{align}
In {\it step 1} of our procedure, for now we choose the simplest $H^0$, where $$H^0_{ij} = \bra{i} H^0 \ket{j} = t < 0$$ for the nearest-neighbor orbitals $i$ and $j$ (indicated by black lines in Fig.~\ref{fig:mod}), and $0$ for any other pair of orbitals. 

Now we move to the next step of our procedure. At first it is not clear how to construct an edge potential satisfying Eq.~\ref{eq:cont_m}, as Eq.~\ref{eq:cont_m} involves states with a well defined angular momentum $m$. Clearly, eigenvectors of $H^0$ can't have a well-defined angular momentum. While the parent free-electron Hamiltonian, $H^0_{\rm c}$ does have a continuous rotational symmetry, this is no longer the case once we projected $H^0_{\rm c}$ into a finite $N \times N$ square mesh of orbitals, to construct our tight-binding model.  Therefore, before discussing the edge perturbation, in {\it step 2} of our procedure we construct a commutator correction term $H^{\rm comm}$ which ensures that total bulk Hamiltonian,
\begin{align}
H^{\rm bulk} = H^0 + H^{\rm comm},
\label{eq:bulk0comm}
\end{align}
at least approximately commutes with the angular momentum operator, $\tilde{L}(H^{\rm bulk})$.  The straightforward but tedious construction of $H^{\rm comm}$ is given in Appendix~\ref{app:comm_construct}.

The energy spectrum of $H^0$ as a function of $N$, shown in the top panel of Fig.~\ref{fig:dos}, exhibits some regularity by having spikes in the density of states separated by $\Delta \sim 1/N$.  However, the number of states in between spikes is not strictly zero, and these states don't follow a clear pattern as a function of increasing $N$. Including the commutator correction term $H^{\rm comm}$ in our Hamiltonian partially restores the continuous rotational symmetry of the Hamiltonian. In the bottom three panels, we show how the incremental addition of $H^{\rm comm}$ to $H^0$ redistributes the states in Fig.~\ref{fig:dos}. In the bottom panel, there are now small gaps in the spectrum (scaling as $\Delta \sim 1/N$) and the states follow a clear pattern as a function of $N$. We find that placing a Fermi level $E_{\rm F}$ within one of these gaps has the additional benefit of stabilizing the finite-size effects in our calculations. Related finite-size effects for Landau diamagnetism have also been reported in Refs.~\onlinecite{ruitenbeek1993review, ruitenbeek1991size, gurevich1997orbital, aldea2003, goldstein2004orbital}.

{\it Step 3:} we now construct edge perturbation 
\begin{align}
V^{\rm edge}_{ij} = - \frac{e B}{2 m_{\rm e}} S_{ij} \tilde{L}_{ij} (H^0).
\label{eq:V_edge}
\end{align}
This term introduces complex phases to the hopping elements on the edge of the model.  Panel (a) of Fig.~\ref{fig:mod} shows a sketch of the alternating effective fluxes applied to the edge of the sample by $V^{\rm edge}$. Sketch of the model for varying $N$ is shown in the supplement.

The $S_{ij}$ term in Eq.~\ref{eq:V_edge} ensures that the perturbing potential $V^{\rm edge}$ is zero in the bulk and non-zero only on the edges. We set $S_{ij} = 0$ when orbitals $i$ and $j$ reside in the interior of the sample. When orbitals $i$ and $j$ are on the edge of the model, we set $S_{ij}$ to a non-zero value.  As specified in Appendix~\ref{app:form_sij}, the non-zero values of $S_{ij}$ are scaling with system size as $1/N$.  This scaling ensures that the complex phase acquired by an electron traversing a closed loop around the edge plaquette (flux) is nearly independent of $N$ and its location along the edge.  Our choice of $S_{ij}$ also ensures that the total flux through the entire sample is zero. Without including $S_{ij}$ in $V^{\rm edge}$, the resulting $V^{\rm edge}_{ij}$ would represent an approximate interaction term of the orbital magnetic moment with a spatially uniform external magnetic field $B$, as in the study of Landau diamagnetism.  Trivially, the matrix element of such a perturbation is proportional to $m$, as in Eq.~\ref{eq:cont_m}. (We note that the complex phases of hopping elements as in $V^{\rm edge}$ are generically present in any magnetic material due to spin-orbit interaction, and they do not require application of external magnetic field.) 

{\it Step 4:} diagonalizing our full Hamiltonian, which includes both bulk and edge contribution,
\begin{align}
\left( H^{\rm bulk} + V^{\rm edge} \right) \ket{\psi_n} = E_n \ket{\psi_n}
\label{eq:schrod}
\end{align}
or, equivalently
\begin{align}
\left( H^0 + H^{\rm comm} + V^{\rm edge} \right) \ket{\psi_n} = E_n \ket{\psi_n}
\end{align}
we obtain a set of eigenstates $\ket{\psi_n}$. The largest model we used has $N=100$, corresponding to a system with 10,000 orbitals. We use even $N$'s, although odd $N$'s yields qualitatively similar results with slightly different chemical potential. We set the Fermi level $E_{\rm F}$ to $-2.55 \, |t|$, placing it within a small energy gap $\Delta$ in the spectrum. 

{\it Step 5:} the magnetic dipole moment we compute as
\begin{align}
m_{\rm dip} &= \frac{e}{2 m_{\rm e}} \sum_n \bra{\psi_n} \tilde{L}(H) \ket{\psi_n} f_n.
\label{eq:morb}
\end{align}
Here $f_n$ is the Fermi-Dirac distribution with effective smearing of electron occupation by $k_{\rm B} T$. 

\section{Results and discussion}\label{sec:results}

\begin{figure}[!t]
\centering
\includegraphics{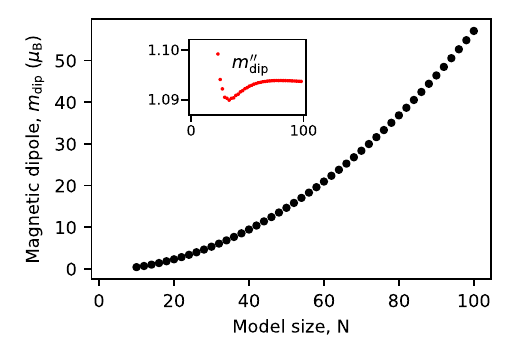}
\caption{\label{fig:mag}  
Changing order $N$ terms in our two-dimensional model induces $N^2$ change in the computed magnetic dipole $m_{\rm dip}$. Here, the temperature $k_{\rm B} T$ in the Fermi-Dirac distribution is set to $0$ and the model is in the microscopic regime. $B$ is chosen so that $a^2 B = 0.2 \hbar/e$.  Fermi level $E_{\rm F}$ is set to $-2.55 \, |t|$ so that the electron density is $\approx 0.12 /a^{2}$. The parameters $t$ and $a$ are set so that the effective mass at low doping is the same as the free electron mass. The inset shows that the second derivative of $m_{\rm dip}$ with respect to $N$ (scaled by $10^2$) is constant.}
\end{figure}

Figure~\ref{fig:mag} shows the calculated $m_{\rm dip}$ as a function of $N$.  The computed $m_{\rm dip}$ is clearly extensive for our two-dimensional model, as it scales nearly perfectly as $N^2$. Clearly, stacking our two-dimensional model to create a three-dimensional solid would result in a $\sim N^3$ scaling of the magnetic moment due to perturbing $\sim N^2$ atoms on the surface.

\begin{figure}[!h]
\centering
\includegraphics{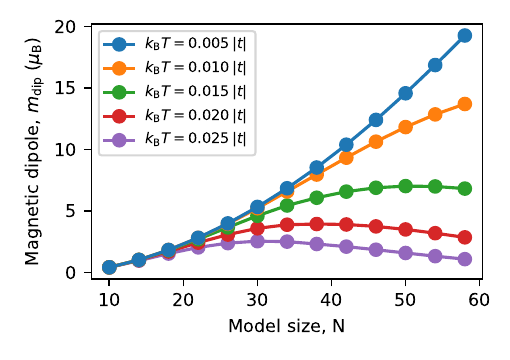}
\caption{\label{fig:mag_temp} Dependence of $m_{\rm dip}$ on $N$ for different values of temperature $T$.   As can be seen from the figure, $m_{\rm dip}$ scales as $N^2$ as long as $k_{\rm B} T$ 
is less than $\approx \frac{0.6 |t|}{N} \approx 0.2 \Delta$.
}
\end{figure}

We find numerically that the extensive scaling persists only when
\begin{align}
 k_{\rm B} T \lesssim 0.2 \, \Delta \approx 0.6 \, \frac{|t|}{N}.
\label{eq:condition}
\end{align}
We show $m_{\rm dip}$ as a function of $N$ at various temperatures in Fig.~\ref{fig:mag_temp}. Since $|t| \sim v_{\rm F}$ and $N \sim L$ clearly Eq.~\ref{eq:condition} is equivalent to the definition of the mesoscopic regime given by Eq.~\ref{eq:meso}.  In other words, $N^2$ scaling of $m_{\rm dip}$ in our two-dimensional model persists only in the mesoscopic regime. (Dependence of $\Delta$ on $|t|$ and $N$ is given in the supplement.)

\begin{figure*}[!t]
\centering
\includegraphics{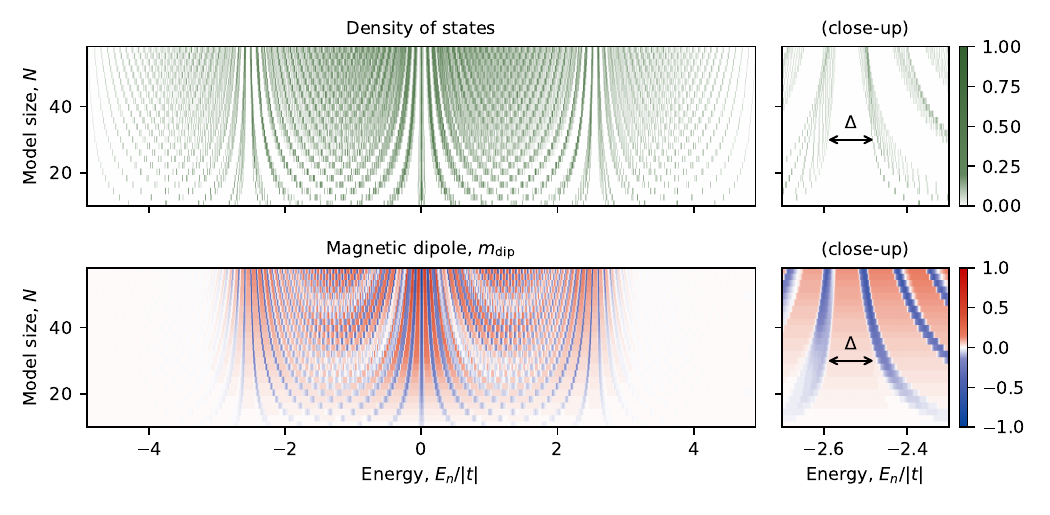}
\caption{\label{fig:vary} 
Top panels show density of states as a function of energy ($E_n$) and $N$. The close-up of the region close to $E_{\rm F} \approx -2.55 \, |t|$ is shown on the right-hand side.  The bottom panels show the magnetic dipole $m_{\rm dip}$ as a function of Fermi level $E_{\rm F}$ and $N$.  The red and blue regions indicate positive and negative $m_{\rm dip}$.}
\end{figure*}

Furthermore, we find that $m_{\rm dip}$ can be fitted well to the following functional form, either in the macroscopic or the mesoscopic regime,
\begin{align}
m_{\rm dip} \sim \frac{N^2}{1 + \exp\left[ 3.8 \displaystyle\frac{k_{\rm B} T}{|t|} \left( N - 0.6 \displaystyle\frac{|t|}{k_{\rm B} T} \right) \right]}.
\label{eq:fit_form}
\end{align}
From this functional form it is clear that $m_{\rm dip} \sim N^2$ in the mesoscopic regime. More precisely, the following mesoscopic limit
\begin{align}
\lim_{N \rightarrow \infty} \lim_{T \rightarrow 0^+} \frac{m_{\rm dip}}{N^2}
\neq 0
\label{eq:limit1}
\end{align}
is non-zero.  In other words, the $N^2$ scaling of the magnetic moment continues for all $N$, as long as the temperature is small enough. Such a limit is shown with a horizontal arrow in the blue region in Fig.~\ref{fig:phase}. On the other hand, if we swap the order of limits, the resulting macroscopic limit
\begin{align}
\lim_{T \rightarrow 0^+} \lim_{N \rightarrow \infty} \frac{ m_{\rm dip}}{N^2}
=0
\label{eq:limit2}
\end{align} 
is now zero. In other words, for any fixed small positive $T$ there is an $N$ beyond which the magnetic dipole no longer scales as $N^2$. Such a limit is indicated with a vertical arrow in the red region in Fig.~\ref{fig:phase}.

Results for orbital magnetic moment at Fermi levels other than $-2.55 \, |t|$ are shown in Fig.~\ref{fig:vary}.  The close-up of the region near $-2.55 \, |t|$ is shown in the right-hand panels of Fig.~\ref{fig:vary}.

\subsection{Consistency checks}\label{subsec:checks}

To ensure that the $m_{\rm dip} \sim N^2$ scaling indeed results from the edge modification, and not from some artifact of the setup, we performed the following numerical consistency checks on our model calculation.

\begin{enumerate}
\item We confirmed numerically that the number of occupied electrons, divided by $N^2$, is a constant as $N \rightarrow \infty$.  Therefore, the order $N^2$ changes in the magnetic dipole in our calculation are not due to variations in the fraction of occupied electronic states.
\item After constructing $H$ from  $H^0$, commutator correction term, and edge perturbation $V^{\rm edge}$, we confirmed that the only $H_{ij}$ terms with an imaginary part  are near the edge of the sample.  In other words, time-reversal breaking edge contributions ($V^{\rm edge}$) are present only on the edge. Furthermore, if we set imaginary terms of $H_{ij}$ to zero, the magnetic dipole moment is zero, as expected.
\item The largest absolute value of the real part of $H_{ij}$ tends to a constant as $N \rightarrow \infty$. Same is true for the imaginary part of $H_{ij}$. Therefore, the $N^2$ scaling of the edge-induced magnetic dipole is not due to scaling of the edge perturbation $V^{\rm edge}$ itself.
\item $H_{ij}$ is zero for all pairs $(i,j)$ that are not nearest neighbors.  Therefore, $H$ is a local operator, even once we include the commutator correction term, and the edge perturbation.
\end{enumerate}

In the supplement~\cite{supp} we provide explicit numerical values of Hamiltonian matrix elements $H_{ij}$ for different values of $N$, as well as a computer code that diagonalizes Eq.~\ref{eq:schrod}, computes Eq.~\ref{eq:morb}, and performs the above consistency checks on $H_{ij}$.

\subsection{Comparison with electric polarization}\label{subsec:polarization}

Now we compare our findings about orbital magnetic dipole with that of the {\it electric} dipole.  To have a more direct comparison between the two cases we will assume that the tight-binding Hamiltonian $H$ is fixed, independent of the occupation of electronic states.  Therefore, we are not including here the electric field generated by the occupied electronic states.  Clearly, such self-consistency effects would automatically exclude the possibility of having a well-defined dipole moment of a metallic state, as the electronic charge would otherwise redistribute to always ensure zero electric field in the bulk (as in the Faraday cage effect).

Even neglecting the effect of self-consistency, the dipole moment of a metallic system is still not a well defined bulk property.  As shown in Ref.~\cite{KingSmith1993} the dipole moment of metallic system under these circumstances would still depend not only on the bulk, but also on the surface properties. 

Interestingly, the electric dipole $d_{\rm dip}$ is edge sensitive in a metal even in the macroscopic regime, unlike the magnetic dipole $m_{\rm dip}$. Therefore, we can naturally ask why, in the macroscopic regime, the magnetic dipole $m_{\rm dip}$ behaves differently from the electric dipole $d_{\rm dip}$? 

To establish a parallel between the electric and magnetic dipole, it is instructive to construct an edge potential $V'^{\rm edge}$ that changes the bulk {\it electric} dipole, in analogy to how $V^{\rm edge}$ changed the bulk magnetic dipole.  To achieve this, we use the following procedure.
\begin{align}
&{\rm Step \ 1': \ choose \ } H^0.
\notag \\
&{\it Step \ 2': \ ( commutator \ correction \ term \ not \ needed.)}
\notag \\
&{\rm Step \ 3': \ construct \ } V'^{\rm edge} {\rm \ from \ } \tilde{x}.
\notag \\
&{\rm Step \ 4': \ diagonalize \ } H = H^0 + V'^{\rm edge}.
\notag \\
&{\rm Step \ 5': \ compute \ } d_{\rm dip} = e \sum_n \bra{\psi_n} \tilde{x} \ket{\psi_n} f_n.
\notag
\end{align}
In step $1'$, we take the same $H^0$ as before.  Step $2'$ is not needed, as we find that a numerically robust $N^2$ scaling of $d_{\rm dip}$ is present even without commutator correction. This is to be expected, as imposing continuous rotational symmetry has direct relevance for a magnetic dipole, but not for an electric dipole.

The important difference is in step $3'$.  Earlier, in the case of the magnetic dipole, we constructed $V^{\rm edge}$ from the angular momentum operator $\tilde{L}(H^0)$, which induced an effective alternating magnetic flux at the edge.  Now, by analogy, in step $3'$ we construct the edge potential from the position operator, 
\begin{align}
V'^{\rm edge}_{ij} = - e {\cal E} S_i  \tilde{x}_i  \delta_{ij},
\label{eq:vprime}
\end{align}
which induces effective {\it electric} fields on the edge, proportional to ${\cal E}$.

Panel (b) of Fig.~\ref{fig:mod} shows the sketch of the effective electric fields near the edge induced by $V'^{\rm edge}$. (This model for various $N$ is shown in the supplement.) In Eq.~\ref{eq:vprime} we use $S_i$ to ensure that the perturbation potential $V'^{\rm edge}$ is zero in the bulk. \footnote{We set $S_i = 0$ for orbitals $i$ in the bulk, and to a non-zero value, scaling as $\sim 1/N$, on the left and right edges of the model.}

\begin{figure}[!t]
\centering
\includegraphics{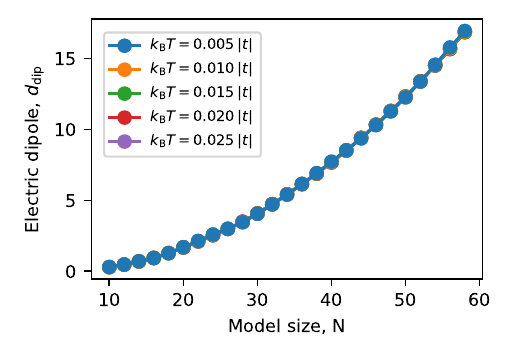}
\caption{\label{fig:pol_temp} Dependence of the {\it electric} dipole $d_{\rm dip}$ on $N$ for different values of temperature $T$ due to edge modification $V'^{\rm edge}$. Here $d_{\rm dip}$ scales as $N^2$ for any $k_{\rm B} T$.}
\end{figure}

\begin{figure*}[!t]
\centering
\includegraphics{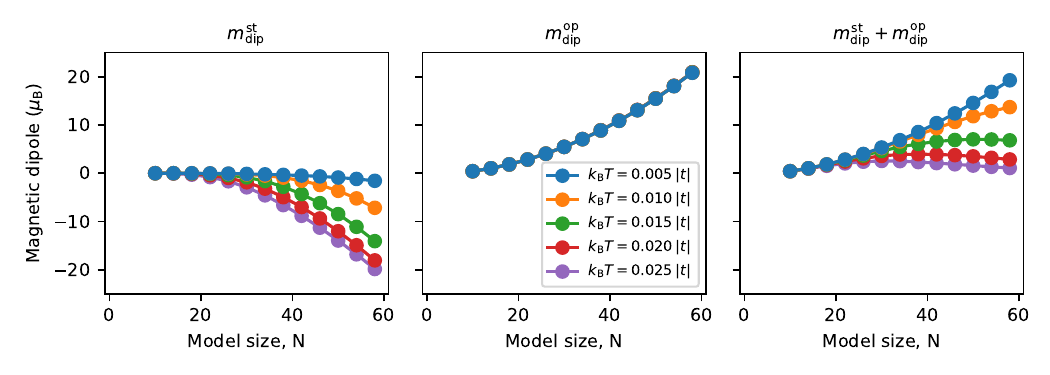}
\caption{\label{fig:parts} Two contributions to $m_{\rm dip}$ cancel each other in the macroscopic regime.}
\end{figure*}

In the final step ($5'$) of our procedure, we now compute the expectation value of the {\it electric} dipole moment, $d_{\rm dip} = e \sum_n \bra{\psi_n} \tilde{x} \ket{\psi_n} f_n.$ As shown in the Fig.~\ref{fig:pol_temp}, we find that $d_{\rm dip}$ scales as $\sim N^2$, even in the macroscopic regime, as expected based on Ref.~\cite{KingSmith1993}.

We assign a different behavior of an electrical dipole to that of a magnetic dipole  due to the fact that the magnetic dipole in step $5$ is computed as a trace over operator $\tilde{L}(H)$ which explicitly includes the edge perturbation $V^{\rm edge}$ itself,
\begin{align}
\tilde{L} (H) = \tilde{L} (H^{\rm bulk}) + \tilde{L} (V^{\rm edge}).
\label{eq:l2}
\end{align}
Therefore, the magnetic dipole $m_{\rm dip}$ can be decomposed into two contributions. The first is a partial trace of $\tilde{L} (H^{\rm bulk})$
\begin{align}
m_{\rm dip}^{\rm st} 
&= 
\frac{e}{2 m_{\rm e}} \sum_n \bra{\psi_n} \tilde{L}(H^{\rm bulk}) \ket{\psi_n} f_n,
\label{eq:morb_st}
\end{align}
and it arises from changes to the electron state (wavefunction and energy) due to edge perturbation $V^{\rm edge}$.
The second term is a partial trace of $\tilde{L} (V^{\rm edge})$
\begin{align}
m_{\rm dip}^{\rm op} 
&= 
\frac{e}{2 m_{\rm e}} \sum_n \bra{\psi_n} \tilde{L}(V^{\rm edge}) \ket{\psi_n} f_n.
\label{eq:morb_op}
\end{align}
and it originates from the change in the angular momentum operator by inclusion of perturbation $V^{\rm edge}$ in the total Hamiltonian.  This term, in the lowest order of perturbation theory, can be computed already from the unperturbed electron wavefunction and energy.

On the contrary, the electric dipole is calculated in step $5'$ as a trace over the position operator $\tilde{x}$, which clearly {\it does not} depend on the edge perturbation $V'^{\rm edge}$.  Therefore, the electric dipole is induced in the model only by changes in the electron wavefunction and energy (analogous to $m_{\rm dip}^{\rm st}$).  In the case of the electric dipole, there are no terms analogous to $m_{\rm dip}^{\rm op}$.

Interestingly, we find that both $m_{\rm dip}^{\rm st}$ and $m_{\rm dip}^{\rm op}$, shown in left and middle panel of Fig.~\ref{fig:parts}, are extensive in the macroscopic regime, on their own. However, in the macroscopic regime, these two terms cancel each other exactly, resulting in a non-extensive magnetic dipole in the macroscopic regime (shown in the right panel of Fig.~\ref{fig:parts}). In contrast, in the case of the electric dipole, there is only one contribution (the one coming from changes in the electron's state), so there is no cancelation, and the electric dipole remains edge sensitive in the macroscopic regime.

\subsection{Chern insulator}\label{subsec:chern}
While the focus of our work has been on topologically trivial materials, the general procedure described here applies to any bulk $H^0$ Hamiltonian, including topological insulators. An interesting case is the Haldane model in a topologically non-trivial insulator phase with a non-zero Chern number.\cite{Haldane1988} Here, repeating our five-step procedure, we find that even when the Fermi level is within the bulk gap and crosses the topologically protected edge states, the induced magnetic dipole $m_{\rm dip}$ is again extensive, and scales as $N^2$. This is numerically robust even without including the commutator correction term $H^{\rm comm}$. We will address the case of Chern insulator in more detail in future work.

\section{Conclusions and outlook}\label{sec:conclusions}
In this work, we showed with a simple model that the orbital magnetization of a metal is not a bulk property in the mesoscopic regime. Instead, one can modify the metal's surface to induce an extensive change in magnetic moment. However, in the macroscopic regime, the orbital magnetization is a bulk property, consistent with previous work.  Therefore, taking the limit of sample size to infinity, the orbital magnetization either is or isn't a bulk property depending on how one takes the limit of infinite sample size.  If the temperature is first taken to zero and the sample size is afterwards taken to infinity, the orbital magnetization is not a bulk property.  On the other hand, if the sample size is first taken to infinity and then the temperature is taken to zero, the orbital magnetization is a bulk property.  These two limits are sketched with horizontal and vertical arrows in Fig.~\ref{fig:phase}.

Our work focuses on the simplest case of a bulk metal, described by a square lattice with first-neighbor hoppings ($H^0$).  Therefore, we expect that a similar phenomenology of magnetic orbital moment will occur rather generally for a more realistic models of a metal.  Similarly, the complex phases on the edge, as in our $V^{\rm edge}$, are generically present in any magnetic material, due to spin-orbit interaction.  Therefore, we expect that the effects discussed in this work could be generically observed in core-shell nanoparticles or nanoparticle assemblies with nonmagnetic core and magnetic shell.  The synthesis of these nanoparticles has been reported in Ref.~\cite{guo2011} for Ag/Ni and in Ref.~\cite{felix2017} for Au/Fe$_3$O$_4$.

Furthermore, if one considers magnetic nanoparticles, one might find that differently terminated surfaces result in different overall, extensive, magnetic dipole, even without adding any additional layers of atoms on the surface.  Clearly, even different surface terminations (for example, surfaces $(100)$ versus $(110)$ versus $(111)$) will induce different surface potentials.

\bigskip
\acknowledgments{This work was supported by the NSF DMR-1848074 grant.  We acknowledge discussions with R.~Wilson and L.~Vuong on inverse Faraday effect as these discussions have motivated our work.}

\appendix
\section{Construction of $H^{\rm comm}$}\label{app:comm_construct}

Given a Hamiltonian $H^0$, we wish to construct a commutator correction term $H^{\rm comm}$ such that
\begin{align}
H^{\rm bulk} = H^0 + H^{\rm comm}
\end{align}
from Eq.~\ref{eq:bulk0comm} approximately commutes with the corresponding angular momentum operator, $\tilde{L}(H^{\rm bulk})$,
\begin{align}
\left[ H^{\rm bulk} , \tilde{L}(H^{\rm bulk}) \right] \approx 0.
\end{align}
Inserting the definition of $H^{\rm bulk}$ from above, and using the linearity of $\tilde{L}$, we obtain
\begin{align}
\left[ H^0 + H^{\rm comm} , \tilde{L}(H^0) + \tilde{L}(H^{\rm comm}) \right] \approx 0.
\end{align}
Expanding the second commutator gives us
\begin{align}
\left[ H^0 + H^{\rm comm}, \tilde{L} \left( H^0 \right) \right]
+ 
\left[ H^0 , \tilde{L} \left( H^{\rm comm} \right) \right] + \nonumber\\
+ \left[ H^{\rm comm}  , \tilde{L}(H^{\rm comm}) \right] \approx 0.
\end{align}

If we keep only the lowest order in $H^{\rm comm}$, neglecting the last term that is quadratic in $H^{\rm comm}$, we are left with the following,
\begin{align}\label{eq:co}
\left[ H^0 + H^{\rm comm}, \tilde{L} \left( H^0 \right) \right]
+
\left[ H^0 , \tilde{L} \left( H^{\rm comm} \right) \right]
\approx 0.
\end{align}
The unknown matrix $H^{\rm comm}_{ij}$ is generally the $N^2 \times N^2 = N^4$ matrix.  Therefore, Eq.~\ref{eq:co} is a system of $N^4$ linear equations with $N^4$ unknowns.

However, we can further restrict $H^{\rm comm}_{ij}$ to zero for distant orbitals $i$ and $j$, making $H^{\rm comm}$ a local operator. This restriction results in a system of only $\sim N^2$ equations.  These equations can be solved using least-square methods.  We perform such a minimization of the left-hand side of Eq.~\ref{eq:co} while varying the system size $N$.  Our approach produces a purely real $H^{\rm comm}$ that only includes the first-nearest neighbors. The maximum value of $|H^{\rm comm}_{ij}|$ is $0.5 |t|$ independently of $N$. The operator $H^{\rm comm}_{ij}$ breaks periodicity in the bulk of the sample and resembles the functional form of a parabolic well. The approximate form of $H^{\rm comm}$ is provided in the following section.  This form was obtained by fitting the results of our procedure for low $N$.  

\section{Approximate form of $H^{\rm comm}_{ij}$}

The coordinate of orbital $i$ is $(x_i, y_i)$, as discussed in Sec.~\ref{sec:construction}.  The allowed values of $x_i$ and $y_i$ are $$0, a, 2a, \ldots, (N-1) a. $$  Now let us introduce the following useful notation,
\begin{align}
d_i^x & = \min \left[ x_i,  (N - 1) a - x_i \right]
,\\ 
d_i^y & = \min \left[ y_i,  (N - 1) a - y_i \right].
\end{align}
The quantities $d_i^x$ and $d_i^y$ measure the distance along the $x$ or $y$ axis to the closest edge (either along $x$ or $y$) of the sample.  Next, we define a similar measure of distance for a pair of points $i$ and $j$,
\begin{align}
d_{ij}^x = \frac{1}{2} \left( d_i^x + d_j^x \right)
,\ \ \ \ 
d_{ij}^y = \frac{1}{2} \left( d_i^y + d_j^y \right)
\end{align}
With this notation, we can now give the approximate form of $H^{\rm comm}_{ij}$.  This form was obtained by first explicitly solving for small $N$ the linear system of equations Eq.~\ref{eq:co}.  Subsequently, we fit the obtained $H^{\rm comm}_{ij}$ to a simple function that can then be easily evaluated for any $N$ without the need to solve the Eq.~\ref{eq:co}.  To give a fitted approximate form of $H^{\rm comm}$ we first define,
\begin{align}
h^{\rm min}_{ij} = \min (d^x_{ij}, d^y_{ij})
,\ \ \ \ 
h^{\rm max}_{ij} = \max (d^x_{ij}, d^y_{ij})
\end{align}
Now we set $H^{\rm comm}_{ij} = 0$ for all $(i,j)$ that are not nearest neighbors.  For nearest neighboring $(i,j)$ we set
\begin{align}
H^{\rm comm}_{ij} &\approx l\left( \frac{h^{\rm max}_{ij}}{N a}\right) (-t)
\end{align}
if $h^{\rm min}_{ij}/a$ is an integer, and 
\begin{align}
H^{\rm comm}_{ij} &\approx l\left( \frac{h^{\rm min}_{ij}}{N a}\right) (-t)
\end{align}
if $h^{\rm min}_{ij}/a$ is not an integer. The function $l(z)$ is defined as $$l(z) = 3 z^2 - 3 z + 1/2.$$

\section{Form of $S_{ij}$}\label{app:form_sij}

The object $S_{ij}$ used in Eq.~\ref{eq:V_edge} 
needs to be zero in the interior and nonzero positive on the edge of the sample.  While there are many $S_{ij}$ that could be used to give the same qualitative result, in this work we report results for a specific choice of $S_{ij}$.  First, we define
\begin{align}
D_i = \min (d^x_i, d^y_i) , \ \ \ \
D_{ij} = \frac{1}{2} \left( D_i + D_j \right).
\end{align}
Therefore, $D_i$ is the distance to the closest edge of the sample, regardless of whether the edge of the sample is on the left, right, top, or bottom side.  Then our $S_{ij}$ is
\begin{align}
S_{ij} = \frac{1}{N} S \left( \frac{D_{ij}}{w} \right).
\end{align}
The function $S(z)$ is defined as,
\begin{align}
S(z) = 16 z^2 (1-z)^2,
\label{eq:func_s}
\end{align}
when $0 < z < 1$ and $S(z) = 0$ otherwise.  The function $S(z)$ has a maximum value of 1 obtained at $z = 1/2$.  Since $S(z)=0$ for $z>1$, this guarantees that $S_{ij} = 0$ whenever $D_{ij} > w$.  All of our calculations are done with $w = 2a$, so that $S_{ij}$ is nonzero only in the two cells closest to the edge.

\begin{figure}[!t]
\centering
\includegraphics{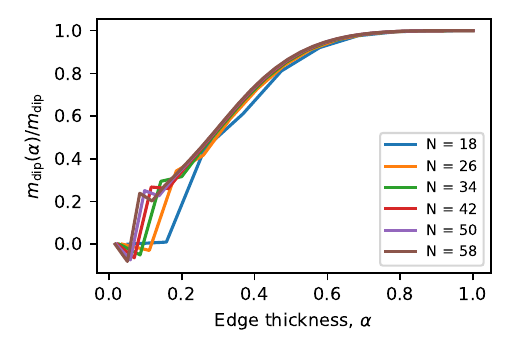}
\caption{\label{fig:alpha}
 Quantity $m_{\rm dip} (\alpha)$ measures contribution of edge to $m_{\rm dip}$.  Thickness of the edge region is parameterized with $\alpha$, as defined in Appendix~\ref{app:edge_contrib}.  About half of the $m_{\rm dip}$ is recovered when $\alpha \approx 0.3$, regardless of $N$.}
\end{figure}

\section{Edge contribution to $m_{\rm dip}$}\label{app:edge_contrib}

The total magnetic dipole of a sample at $k_{\rm B}T = 0$ we compute as $m_{\rm dip} = \frac{e}{2 m_{\rm e}} \sum_n^{\rm occ} \bra{\psi_n} \tilde{L}(H) \ket{\psi_n}$.
Now we wish to get the edge contribution to $m_{\rm dip}$.  For that purpose we define operator $\mathcal{E} (\alpha)$ to project into edge orbitals only,
\begin{align}
\mathcal{E} (\alpha) = \sum_{D_i \leq \alpha \left(\frac{N}{2} - 1 \right) a} \ket{i} \bra{i}.
\end{align}
The thickness of the edge region is $\alpha \left(\frac{N}{2} - 1 \right) a$.  The parameter $\alpha$ is a number between $0$ and $1$.  If $\alpha$ is a small positive number, then only a few sites adjacent to the edge are included in $\mathcal{E}$.  If $\alpha = 1$, the effective edge region is so thick that $\mathcal{E}$ includes the entire sample. 

If we further define $\bar{\mathcal{E}} = 1 - \mathcal{E}$ to be a projector into interior orbitals (those that are not on the edge), then by insertion of unity we have
$$
m_{\rm dip} = \frac{e}{2 m_{\rm e}} \sum_n^{\rm occ} 
\bra{\psi_n} 
(\mathcal{E} + \bar{\mathcal{E}})
\tilde{L}(H) 
(\mathcal{E} + \bar{\mathcal{E}})
\ket{\psi_n}.
$$
Expanding the product we get
\begin{widetext}
\begin{align}
m_{\rm dip} = 
\frac{e}{2 m_{\rm e}} \sum_n^{\rm occ} 
\bigg[ 
\bra{\psi_n} 
\mathcal{E}
\tilde{L}(H) 
\mathcal{E}
\ket{\psi_n} 
+
\bra{\psi_n} 
\bar{\mathcal{E}}
\tilde{L}(H) 
\mathcal{E}
\ket{\psi_n}
+
\bra{\psi_n} 
\mathcal{E} 
\tilde{L}(H) 
\bar{\mathcal{E}}
\ket{\psi_n}
+
\bra{\psi_n} 
\bar{\mathcal{E}}
\tilde{L}(H) 
\bar{\mathcal{E}}
\ket{\psi_n}
\bigg].
\end{align}
\end{widetext}
We find numerically that the cross-terms (second and third term above) are small in comparison to the first term for most $\alpha$.  We can then use the first term,
\begin{align}
m_{\rm dip} (\alpha) &= \frac{e}{2 m_{\rm e}} \sum_n^{\rm occ} \bra{\psi_n} \mathcal{E} (\alpha) \tilde{L}(H) \mathcal{E} (\alpha) \ket{\psi_n},
\end{align}
as a measure of contribution of the edge to the magnetic dipole $m_{\rm dip}$. We show $m_{\rm dip} (\alpha)$ for various $N$ in Fig.~\ref{fig:alpha}.  As can be seen from the figure, about half of the extensive magnetic dipole is recovered when parameter $\alpha$ is close to 0.3.

\bibliography{pap}

\foreach \x in {1,2,3,4}
{%
\clearpage
\includepdf[pages=\x]{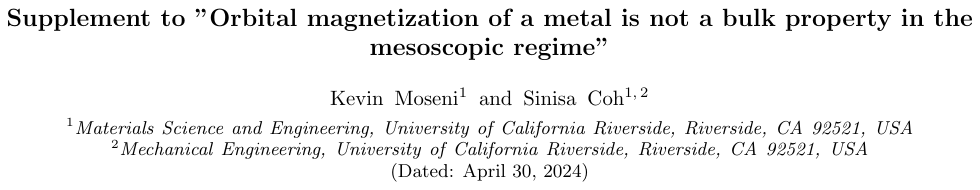}
}

\end{document}